\documentclass[11pt]{article}
\usepackage{latexsym}
\newcommand{\beq}{\begin{equation}}
\newcommand{\eeq}{\end{equation}}
\newcommand{\beqs}{\begin{eqnarray}}
\newcommand{\eeqs}{\end{eqnarray}}
\newcommand{\twist}[2]{\raisebox{-2mm}{$\stackrel{#1\;\Box}
          {\;\;\;\scriptstyle{#2}}$}}
\begin{document}
\begin{titlepage}
\vskip 1.5cm
\begin{center}
{\LARGE \bf D-brane probes on $G_2$ Orbifolds}\\ 
\vfill {\Large G. Ferretti, P. Salomonson and D. Tsimpis} \\
\vfill { \sl Institute for Theoretical Physics - G\"oteborg University and \\
Chalmers University of Technology, 412 96 G\"oteborg, Sweden}
\end{center}
\vfill
\begin{abstract}
\noindent We consider type IIB string 
theory on a seven dimensional orbifold with 
holonomy in $G_2$. The motivation is to use D1-branes as probes of the
geometry. The low energy theory on the D1-brane is a sigma-model 
with two real supercharges (${\cal N} = (1,1)$ in two dimensional language). 
We study in detail the closed and open string sectors and propose a
coupling of the twisted fields to the brane that 
modifies the vacuum moduli space so that the
singularity at the origin is removed. 
Instead of coming from D-terms, which are not present here,
the modification comes from a ``twisted'' mass term for the seven scalar
multiplets on the brane. The proposed mechanism involves a
generalization of the moment map.

\vskip 0.2 cm 
\end{abstract}
\end{titlepage}

\section{Introduction}

Recently, there has been a focus of attention on manifolds with $G_2$ 
holonomy \cite{joyce}. 
The physical motivation, (see, e.g. \cite{Acharya:1999pm},
\cite{Acharya:2000gb}, \cite{Atiyah:2000zz}, \cite{Atiyah:2001qf})
is very clear and very strong -- M-theory
on such manifolds gives rise to ${\cal N}=1$ supersymmetry 
in $3+1$~dimensions. 
Phenomenologically interesting situations correspond
to regions in the moduli space where the
$G_2$ manifold develops singularities
and non-perturbative effects arise. 
Investigations of 
the local physics near such singularities
has already unearthed many new interesting phenomena. For a partial list of 
$G_2$-related literature, see
\cite{Shatashvili:zw} - \cite{Acharya:2001hq}. 

In this paper, we will consider type II string theory
on orbifolds \cite{Dixon:1985jw}, \cite{Dixon:1986jc},
\cite{Dixon:1987qv}
whose holonomy is in $G_2$.
Our motivation is
to provide a setup where it is possible to apply
the brane-probe techniques
initiated by Douglas and Moore \cite{Douglas:1996sw} and
expanded in \cite{Johnson:1997py} - \cite{Greene:1999rr}. 
The idea is that
the use of non-perturbative objects as probes
unravels aspects of the geometry that depart from
the classical picture. 
The effective spacetime geometry as `seen' by the D-brane
probes is the moduli space of vacua of the low-energy theory
on their world-volume. The process by which 
D-branes resolve orbifold singularities is by
the couplings of their 
world-volume theory to twisted bulk fields. These additional 
bulk couplings smooth-out the vacuum moduli space.
In \cite{Douglas:1996sw} 
this was applied to $\mathbf{C}^2/\mathbf{Z}_n$ orbifold singularities
and the hyperK\"ahler quotient
construction \cite{Lindstrom:rt}, \cite{Hitchin:1987ea}, 
\cite{Kronheimer:1989zs} (see also \cite{hit})
was reproduced from a `brany' point of view.

The fact that the 
amount of unbroken supersymmetry in our case is 1/4
of that in \cite{Douglas:1996sw}, makes our setup rather different.
From the technical point of view, the low number
of supercharges does not allow for D-terms --
which in \cite{Douglas:1996sw} were responsible for the resolution
of the singularity!
Instead of the D-terms, which are not present here,
we propose that the singularity gets removed
due to a ``twisted'' mass term for the seven scalar
multiplets on the brane. 
Our analysis is valid for lengths $r$
in the regime $l_P << r << \sqrt{\alpha^\prime}$.

This paper is organized as follows:
In section 2 we introduce our $G_2$ orbifold and 
in section 3 we discuss
the closed string spectrum for type II string 
theory on it, paying particular attention to the massless modes
and the partition function. In section 4 we discuss the spectrum of 
open strings ending
on D-branes and the corresponding gauge theory arising from their interactions.
In section 5 we combine the results and propose a
coupling of the twisted fields to the brane that removes
the singularity at the origin. 
As a byproduct of the construction we propose a 
generalization of the moment map.
We discuss the significance and limitations
of our results in the last section.

\section{The orbifold}

To our knowledge, no classification of the possible discrete subgroups of 
$G_2$ is known. In this paper we will be concerned with what is arguably 
the simplest such orbifold: 
$\Gamma = {\bf Z}_2 \times {\bf Z}_2 \times {\bf Z}_2$ where the three 
generators $\alpha$, $\beta$ and $\gamma$ act on the last seven coordinates
$X^3, \cdots X^9$ as follows:

\bigskip
\begin{center}
  \begin{tabular}{c||c|c|c|c|c|c|c|}
    & $X^3$ & $X^4$  & $X^5$  & $X^6$  & $X^7$  & $X^8$  & $X^9$ \\
    \hline
    \hline
    $\alpha$ :& $-X^3$ & $-X^4$ & $-X^5$ & $-X^6$  &
               $X^7$  & $X^8$  & $X^9$ \\
    \hline
    $\beta$ :& $-X^3$ & $-X^4$ & $X^5$ & $X^6$  &
               $-X^7$  & $-X^8$  & $X^9$ \\
    \hline
    $\gamma$ :& $-X^3$  & $X^4$  & $-X^5$  & $X^6$  & 
                $-X^7$  & $X^8$  & $-X^9$ \\
   \end{tabular}
\end{center}
\bigskip

Since we are interested only in the local physics 
near the singularity, we take
our orbifold to be non-compact, i.e. of the form ${\bf R}^7/\Gamma$.
By doing this we are implicitly assuming that
we are focusing our attention to an open neighborhood
$U$ of a singular point in a {\it compact}
manifold $M$, such that $U$ is isometric to 
${\bf R}^7/\Gamma$.

An important question, 
which we are not able to answer here, is whether $M$ can be taken to be a 
$G_2$ manifold.
There are several known compact $G_2$ manifolds,
constructed by Joyce in \cite{joycetwo},
(For an occurrence of such orbifolds in
the physics literature see \cite{Acharya:1999pm})
by desingularizing 
compact orbifolds  of the type $T^7/({\bf Z}_2)^3$.  
The set of singular points in Joyce's examples
consists of disjoint unions of 
singularities of the type $T^3\times \mathbf{C}^2/{\bf Z}_2$,
and are thus pretty well behaved.
In particular, each singular patch can be desingularized
much in the same way $\mathbf{C}^2/{\bf Z}_2$ can be desingularized
to give the Eguchi-Hanson space \cite{Eguchi:1978xp}.
 Compared to the examples given in \cite{joycetwo}, our orbifold differs
by the fact that none of the seven coordinates is compact and there is no
associated ``shift'' in those coordinates. 
The  singularity of our non-compact orbifold
is a `bad' one comparatively,
in the sense that it cannot be smoothed-out
by the method of \cite{joycetwo}.

It is easy to see that 
$\Gamma\subset G_2$ but $\Gamma \not\subset SU(3)$. 
Recall that geometrically $G_2$ 
can be thought of as being
generated by simultaneous rotations in two orthogonal
planes in $\mathbf{R}^7$. (This is actually one way to 
show that $G_2$ is embedded in $Spin(7)$). 
Here we are dealing only with rotations by $\pi$, so we
see that a group generated by seven-dimensional diagonal matrices with
entries equal to $\pm 1$ on the diagonal, will be in $G_2$ iff all the
elements other than the identity
have exactly four negative entries (thus rotating two orthogonal planes 
by $\pi$). This property is indeed satisfied by the group generated by
the $Spin(7)$ matrices
\beqs
R(\alpha)&=& \mathrm{diag}(-1,-1,-1,-1,1,1,1) \nonumber\\
R(\beta)&=& \mathrm{diag}(-1,-1,1,1,-1,-1,1) \nonumber\\
R(\gamma)&=& \mathrm{diag}(-1,1,-1,1,-1,1,-1). 
\eeqs

However, trying to add a generator of type 
$\mathrm{diag}(-1,-1,-1,1,-1,1,1)$ would generate matrices with only two
negative entries, thus taking us outside of $G_2$ to the full $Spin(7)$. 
To check that the orbifold group $\Gamma$ is not a discrete subgroup of
a smaller Lie subgroup of $Spin(7)$ (for instance $SU(3)$) we must check
that it acts non trivially on all the coordinates. For instance, the
subgroup generated by $R(\alpha)$ and $R(\beta)$ alone is in $SU(3)$.

\section{The closed string spectrum}

We begin by looking at the closed string spectrum (no branes) of type II 
string theory on $\mathbf{R}^7/\Gamma$. 
We write the closed string mode expansion as
$X^M(\tau, \sigma) = X^M_R(\tau-\sigma) + X^M_L(\tau+\sigma)$ where, as usual, 
\beqs
    X^M_R(\tau-\sigma)
      &=&\frac{1}{2}q^M + \sqrt{2\alpha^\prime}(\tau-\sigma)\alpha^M_0 +
     i\sqrt{\frac{\alpha^\prime}{2}}\sum_t \frac{\alpha^M_t}{t} 
     e^{-2it(\tau-\sigma)}
     \nonumber\\
    X^M_L(\tau+\sigma)&=&\frac{1}{2}q^M + \sqrt{2\alpha^\prime}(\tau+\sigma)
     \tilde\alpha^M_0 +
     i\sqrt{\frac{\alpha^\prime}{2}}\sum_t \frac{\tilde\alpha^M_t}{t} 
     e^{-2it(\tau+\sigma)}.
\eeqs

For $\mu = 0, 1, 2$ we have the usual closed string boundary conditions
$X^\mu(\tau, 0) = X^\mu(\tau,\pi)$ satisfied for 
$\alpha_0^\mu = \tilde\alpha_0^\mu=\sqrt{\frac{\alpha^\prime}{2}} p^\mu$ and
$t \in {\bf Z} - \{0\}$. 
The other boundary condition, (needed
for $i$ taking the appropriate values within $\{3, \cdots, 9 \}$ in each 
sector), is $X^i(\tau, 0) = - X^i(\tau,\pi)$ and yields 
$\alpha^i_0 = \tilde\alpha^i_0 = 0$, $q^i = 0$ and $t \in {\bf Z} + 1/2$.

The zero point energy in \emph{all} twisted sectors is zero because there are
always four half-odd moded bosonic coordinates. 
Let us consider type IIA for definiteness.
Of the 64 untwisted massless d.o.f. in each of the four GSO sectors of the
superstring ($(NS_+,NS_+)$, $(NS_+,R_-)$, $(R_+,NS_+)$, and $(R_+,R_-)$)
only 8 for each sector survive the orbifold projection. One can see this
by looking at the field description of such d.o.f. -- for instance, in the
$(NS_+,NS_+)$ sector we are left with $g_{\mu\nu}$ and $B_{\mu\nu}$, carrying
no d.o.f. in $d=2+1$ dimensions, the dilaton $\Phi$ and seven more scalars 
from $g_{ii}$, $i=3,\cdots 9$ for a total of eight d.o.f. Similarly, in the
$(R_+,R_-)$ sector we have one d.o.f. from $A_\mu$, dual to a scalar in 
$d=2+1$ dimensions, and seven more scalars from $A_{ijk}$ with the 
appropriate choice of indices.
In type IIB the $(R_+,R_-)$ sector is replaced by $(R_+,R_+)$
and includes one scalar $\chi$  (the axion) and seven
3d scalars $C_{ijkl}^+$.

We can readily see that one-eighth of supersymmetry 
(${\cal N}=2$ in (2+1)-dimensions) is preserved
by the orbifold by noting that only 8 of the 64 fermionic
states survive. Alternatively, it is straightforward to check directly
that there is exactly one spinor of $Spin(7)$ 
invariant under the orbifold action. 

One can obtain the same result by considering the explicit form of the
action of $\alpha$, $\beta$ and $\gamma$ on the states. For instance, in the
$(R_+,R_-)$ sector one needs the action $S(\alpha)$ $S(\beta)$ and
$S(\gamma)$ of the 
generators on the fermionic zero modes\footnote{We denote by $S$ the eight  
dimensional representation of $\Gamma$ acting on spinors, not to 
be confused 
with the seven dimensional one $R$, introduced before, which acts on
vectors.}. Since all three generators represent
a rotation by $\pi$ in two separate planes, it follows that, for instance,
\beq
     S(\alpha)=\exp{i\pi(\Sigma_{34}+\Sigma_{56}})\otimes 
     \exp{i\pi(\tilde\Sigma_{34}+\tilde\Sigma_{56}})=
      \Gamma^3\Gamma^4\Gamma^5\Gamma^6\otimes
      \tilde\Gamma^3\tilde\Gamma^4\tilde\Gamma^5\tilde\Gamma^6
\eeq
when acting on the vacuum state in $|R_+\rangle \otimes |R_-\rangle$.
Similarly
\beqs
    S(\beta) &=&\Gamma^3\Gamma^4\Gamma^7\Gamma^8\otimes
      \tilde\Gamma^3\tilde\Gamma^4\tilde\Gamma^7\tilde\Gamma^8
    \nonumber\\
    S(\gamma) &=&\Gamma^3\Gamma^5\Gamma^7\Gamma^9\otimes
      \tilde\Gamma^3\tilde\Gamma^5\tilde\Gamma^7\tilde\Gamma^9.
\eeqs 
The above matrices commute and thus can be simultaneously diagonalized leaving
invariant 8 out of the 64 d.o.f. in the untwisted $(R_+,R_-)$ sector.

As far as the twisted sectors are concerned, 
for each one of them and for each of the GSO sectors, we have exactly one
degree of freedom. Thus we have a total of 7 twisted $(NS_+,NS_+)$
scalar fields in (2+1)-dimensions (there are 7 twisted sectors), all
corresponding to $B$-field moduli\footnote{In the previous version of
the paper, we had incorrectly identified these scalars with metric
deformations. We thank B.~Acharya for pointing this out to us.}.

One quick way to understand the above counting is to notice that 
each generator taken alone would give us a model
identical to the $\mathbf{C}^2/\mathbf{Z}_2$ orbifold, for which it is well
known that there are four d.o.f. 
in each twisted GSO sector. Keeping only those 
states that are invariant under the action of the other generators reduces 
their number by a quarter. Once again, this can be seen by
looking at the action of the generators on the fermionic zero modes. 
Here, due to the anti-periodicity of some bosonic coordinates, there will be 
fermionic zero modes even in the $(NS_+,NS_+)$ sector. For
instance, in the $\alpha$ sector, we have
\beqs
    S(\alpha) &=&\Gamma^3\Gamma^4\Gamma^5\Gamma^6\otimes
      \tilde\Gamma^3\tilde\Gamma^4\tilde\Gamma^5\tilde\Gamma^6
    \nonumber\\
    S(\beta) &=&\Gamma^3\Gamma^4\otimes
      \tilde\Gamma^3\tilde\Gamma^4
    \nonumber\\
    S(\gamma) &=&\Gamma^3\Gamma^5\otimes
      \tilde\Gamma^3\tilde\Gamma^5.
\eeqs 
These generators commute with each other and reduce 
the number of d.o.f. by a quarter.

This analysis carries over to type IIB virtually unchanged.
The result is again that there is exactly one d.o.f. for each
twisted sector in each one of the GSO sectors.

It is also instructive to look at the partition function of the theory. 
Let us denote by \twist{g}{h} the $8\times 8=64$ sectors of
the orbifold. All the sectors are of course zero 
due to the cancellation of 
bosons against the 
fermions but many of them are ``\emph{trivially zero}'' 
in the sense that 
the bosonic and fermionic parts 
cancel separately. It is fairly easy to check
that the only sectors that are \emph{not} trivially zero are:
\twist{e}{e}, \twist{e}{h}, and \twist{h}{h}
for $h\in\{\alpha, \beta, \gamma, \alpha\beta, \alpha\gamma, \beta\gamma, 
\alpha\beta\gamma \}$, the 7 twisted sectors. 
Moreover,  all \twist{e}{h} contributions are obviously the same
and similarly for the \twist{h}{h}.

The full partition function is thus
\beq
    Z_{\mathbf{R}^7/\Gamma} = 
   \frac{1}{8}\left(\twist{e}{e} + \sum_{h\not=e}\twist{e}{h}+ 
   \sum_{h\not=e}\twist{h}{h}\right) \equiv
   \frac{1}{8}\left(\twist{e}{e} + 7\twist{e}{\alpha}+ 
   7\twist{\alpha}{\alpha}\right).
\eeq
We see immediately that only one eighth of the untwisted fields survives.
As far as the twisted fields go, recall that orbifolding only by, say, 
$\alpha$, we obtain
\beq
    Z_{\mathbf{C}^2/\mathbf{Z}^2} = \frac{1}{2}\left(\twist{e}{e} + 
   \twist{e}{\alpha} + \twist{\alpha}{\alpha}\right).
\eeq
We see that we have $2 \times (7/8)$ as many d.o.f. as in the 
$\mathbf{C}^2/\mathbf{Z}^2$ twisted sector, i.e. $2 \times (7/8) \times 4 = 7$.

Let us try to understand the massless spectrum 
from a geometrical perspective.
Let us denote by $\tilde\omega^I, \,\, I=1\dots 7,$ 
the two-forms on which we expand the $B$-field.
In addition we have 7 (non-normalizable)
$\Gamma-$invariant 
three-forms (and their dual four-forms)
which we denote by $\omega^a, \,\, a=1\dots 7$.
These can be taken to be $dx^i\wedge dx^j\wedge dx^k$
with 
\beq
(ijk)=\{(394),\;(358),\;(367),\;(475),\;(468),\;(569),\;(789)\}.
\eeq
If our orbifold is though of as part
of a compact singular manifold, the $\omega^a$'s will become normalizable. 

The untwisted fields in the NS-NS sector 
are the dilaton $\Phi$ and 7 scalars coming from the 
$\Gamma-$invariant 
metric deformations $\theta_{ii}$.
In the R-R untwisted sector we have the axion $\chi$ and 7
one-forms $c^{a}_{\mu}$ obtained by reducing $C^{+}_{MNKL}$
along $\omega^a$. In the light-cone it is easy to see
that the one-forms are dual to 7 scalars $c^{a}$ which
are obtained by  reducing $C^{+}_{MNKL}$
along ${}^*\omega^a$ --the Hodge star acting on
the internal 7-dimensional space.
In the twisted NS-NS sectors there are 7 scalars
$\phi^I$ $I=1\cdots 7$, 
coming from the twisted $B$-field.
Finally, the R-R twisted sectors contain 7 scalars $c^I$ which
come from expanding  $C_{MN}$ along $\tilde\omega^I$. 
All this is summarized in the following table

\smallskip
\begin{center}
  \begin{tabular}{c||c|c|c|c|c|}
    & NS-NS & R-R  \\
    \hline
    \hline
    untwisted :&  $ \Phi, \theta_{ii}$&
$\chi,  c^{a}  $\\
    \hline
    twisted :& $ \phi^I$  & $ c^I$\\
   \end{tabular}
\end{center}
\bigskip

In the case of K3 surfaces, it is well known
(see \cite{Aspinwall:1996mn} for a nice review)
that sixteen ${\bf C}^2/{\bf Z}_2$ blown-up orbifold singularities 
may be patched together to form the 
Kummer surface --a special kind of K3 surface.
Moreover, Joyce's orbifolds
mentioned in section 2 are just 
$G_2$ analogues of the Kummer surface. 
The reader  may  wonder whether
our seven-dimensional orbifold is to 
any of these `7-dimensional Kummer surfaces'
what ${\bf C}^2/{\bf Z}_2$ is to K3.
However it is not. As
already mentioned, our singularity
is more complicated than the singularities
in Joyce's examples and cannot be
smoothed-out in the same way.

\section{The open string spectrum}

In order to perform the probe analysis of the next section
we need the massless sector of the spectrum of
D1 excitations, where the D1-brane is placed along $\mu=0,1$.
The motion of the brane transverse to the orbifold
is parameterized by the field $X^{\mu},\, \mu=2$, while 
$\{X^{i}, i=3\dots 9\}$ 
parameterize the motion of D1 along the orbifolded directions.
On the brane lives also a gauge field $A_{\mu}, \mu=0,1$ and, of course,
the fermionic degrees of freedom required by supersymmetry.

The low-energy effective theory on the D1 brane world-volume 
is a linear 2d supersymmetric $\sigma-$model. Its supersymmetry
can be determined in the following way. Let $Q_{L,R}$ 
be the supercharges associated to left, right-moving worldsheet
degrees
of freedom of type IIB string theory. The closed-string
sector is invariant under supersymmetry 
transformations of the form $\epsilon_L Q_L+\epsilon_R Q_R$
where $\epsilon_{L,R}$ are Majorana-Weyl 10d spinors,
i.e. they are in the ${\bf 16}_+$ of $Spin(1,9)$.
A D1 brane
along the $\mu=0,1$ directions is invariant under
the subset of the above supersymmetry transformations
which obey in addition $\epsilon_L=\Gamma^{01}\epsilon_R$.
This means that $\epsilon_L$  (say) 
can be expressed in terms of  $\epsilon_R$ leaving
the latter as the only independent supersymmetry parameter.
In other words the D1 breaks half the supersymmetry.
It is useful to decompose $\epsilon_R$ under
$Spin(1,1)\otimes Spin(8) \subset Spin(1,9)$ so that
$\epsilon_R \sim {\bf 8}_{1\over 2}\oplus {\bf 8}_{-{1\over 2}}$,
where the subscripts denote $Spin(1,1)$ weight.
Upon further compactifying 7 of the 8 transverse directions
on a $G_2$ manifold (orbifold) only the singlets of $G_2$ survive as 
unbroken supersymmetry parameters. Noting that the 
${\bf 8}$ of $Spin(8)$ decomposes under $G_2$ into 
${\bf 7}\oplus{\bf 1}$, we conclude that the unbroken
supersymmetry parameter transforms as the
$\epsilon_R \sim {\bf 1}_{1\over 2}\oplus {\bf 1}_{-{1\over 2}}$
of $Spin(1,1)\otimes G_2$. In other words the
theory on the D1 brane will be a linear $(1,1)$ 
$\sigma-$model.

\smallskip

{\it Consistency conditions}

\smallskip

In order to find the massless open string spectrum we follow a well 
known procedure (see, e.g. \cite{Pradisi:1989xd}, \cite{Gimon:1996rq}).
Let $\rho$ be the regular representation\footnote{Yet another representation, 
not to be confused with the previous ones $R$ and $S$.} of $\Gamma$. It is of
course eight-dimensional and it is straightforward to
see that for an appropriate choice of basis it takes the form
\beq
\rho(\alpha)={\bf 1}\otimes {\bf 1}\otimes \sigma^3;\,\,\,\,
\rho(\beta)= {\bf 1}\otimes \sigma^3\otimes {\bf 1};\,\,\,\,
\rho(\gamma)= \sigma^3\otimes {\bf 1}\otimes {\bf 1}.
\eeq

The field $X^{\mu},\, \mu=2$, satisfies
\beq
\rho(\alpha)X^2\rho(\alpha)=\rho(\beta)X^2\rho(\beta)=
\rho(\gamma)X^2\rho(\gamma)=X^2, 
\,\,\,\, (X^2)^{\dagger}=X^2.
\eeq
The solution of the above equations is
\beq
X^2=\pmatrix{  x^2_1 & 0 & 0 & 0 & 0 & 0 & 0 & 0\cr 
 0 & x^2_2 & 0 & 0 & 0 & 0 & 0 & 0\cr 
 0 & 0 &  x^2_3 & 0 & 0 & 0 & 0 & 0\cr 
 0 & 0 & 0 & x^2_4 & 0 & 0 & 0 & 0\cr 
 0 & 0 & 0 & 0 & x^2_5 & 0 & 0 & 0\cr 
 0 & 0 & 0 & 0 & 0 & x^2_6 & 0 & 0\cr 
 0 & 0 & 0 & 0 & 0 & 0 & x^2_7 & 0\cr 
 0 & 0 & 0 & 0 & 0 & 0 & 0 & x^2_8\cr  }
\eeq
where $x^2_1,\dots x^2_8 \in {\bf R}$.

The gauge field satisfies the same condition and is 
also given by a real, diagonal $8\times 8$ matrix, 
implying a $U(1)^8$ gauge symmetry. The
diagonal $U(1)$, which 
describes the center-of-mass of the branes, decouples and
will play no role in the following.
The `real' gauge group is therefore $U(1)^7$.

The fields $X^{3, \cdots,9}$ satisfy
\beqs
\rho(\alpha)X^i\rho(\alpha)=R(\alpha)^i_j X^j,
 \nonumber\\
\rho(\beta)X^i\rho(\beta)=R(\beta)^i_j X^j,
 \nonumber\\
\rho(\gamma)X^i\rho(\gamma)=R(\gamma)^i_j X^j,
 \nonumber\\
 (X^i)^{\dagger}=X^i, \,\,\, i=3\dots 9.
\eeqs
The solution of the above equations can be given 
in terms of 28 complex numbers $x^i,y^i,z^i,w^i, i=3\dots 9$
and reads

{\footnotesize{
\beqs
X^3&=&\pmatrix{  0 & 0 & 0 & 0 & 0 & 0 & 0 & x^3\cr 
               0 & 0 & 0 & 0 & 0 & 0 & y^3 & 0\cr 
               0 & 0 & 0 & 0 & 0 & z^3 & 0 & 0\cr 
               0 & 0 & 0 & 0 & w^3 & 0 & 0 & 0\cr 
               0 & 0 & 0 & \bar{w}^3 & 0 & 0 & 0 & 0\cr 
               0 & 0 & \bar{z}^3 & 0 & 0 & 0 & 0 & 0\cr 
               0 & \bar{y}^3 & 0 & 0 & 0 & 0 & 0 & 0\cr
               \bar{x}^3 & 0 & 0 & 0 & 0 & 0 & 0 & 0\cr },\,\,\,\,\,
X^4=\pmatrix{  0 & 0 & 0 & 0 & 0 & 0 & x^4 & 0\cr 
               0 & 0 & 0 & 0 & 0 & 0 & 0 & y^4\cr 
               0 & 0 & 0 & 0 & z^4 & 0 & 0 & 0\cr 
               0 & 0 & 0 & 0 & 0 & w^4 & 0 & 0\cr 
               0 & 0 & \bar{z}^4 & 0 & 0 & 0 & 0 & 0\cr 
               0 & 0 & 0 & \bar{w}^4 & 0 & 0 & 0 & 0\cr 
               \bar{x}^4 & 0 & 0 & 0 & 0 & 0 & 0 & 0\cr
               0 & \bar{y}^4 & 0 & 0 & 0 & 0 & 0 & 0\cr }
\nonumber\\
X^5&=&\pmatrix{  0 & 0 & 0 & 0 & 0 & x^5 & 0 & 0\cr 
               0 & 0 & 0 & 0 & y^5 & 0 & 0 & 0\cr 
               0 & 0 & 0 & 0 & 0 & 0 & 0 & z^5\cr 
               0 & 0 & 0 & 0 & 0 & 0 & w^5 & 0\cr 
               0 & \bar{y}^5 & 0 & 0 & 0 & 0 & 0 & 0\cr 
               \bar{x}^5 & 0 & 0 & 0 & 0 & 0 & 0 & 0\cr 
               0 & 0 & 0 & \bar{w}^5 & 0 & 0 & 0 & 0\cr 
               0 & 0 & \bar{z}^5 & 0 & 0 & 0 & 0 & 0\cr},\,\,\,\,\,
X^6=\pmatrix{  0 & 0 & 0 & 0 & x^6 & 0 & 0 & 0\cr 
               0 & 0 & 0 & 0 & 0 & y^6 & 0 & 0\cr 
               0 & 0 & 0 & 0 & 0 & 0 & z^6 & 0\cr 
               0 & 0 & 0 & 0 & 0 & 0 & 0 & w^6\cr 
               \bar{x}^6 & 0 & 0 & 0 & 0 & 0 & 0 & 0\cr 
               0 & \bar{y}^6 & 0 & 0 & 0 & 0 & 0 & 0\cr 
               0 & 0 & \bar{z}^6 & 0 & 0 & 0 & 0 & 0\cr 
               0 & 0 & 0 & \bar{w}^6 & 0 & 0 & 0 & 0\cr }
\nonumber\\
X^7&=&\pmatrix{  0 & 0 & 0 & x^7 & 0 & 0 & 0 & 0\cr 
               0 & 0 & y^7 & 0 & 0 & 0 & 0 & 0\cr 
               0 & \bar{y}^7 & 0 & 0 & 0 & 0 & 0 & 0\cr 
               \bar{x}^7 & 0 & 0 & 0 & 0 & 0 & 0 & 0\cr 
               0 & 0 & 0 & 0 & 0 & 0 & 0 & z^7\cr 
               0 & 0 & 0 & 0 & 0 & 0 & w^7 & 0\cr 
               0 & 0 & 0 & 0 & 0 & \bar{w}^7 & 0 & 0\cr 
               0 & 0 & 0 & 0 & \bar{z}^7 & 0 & 0 & 0\cr},\,\,\,\,\,
X^8=\pmatrix{  0 & 0 & x^8 & 0 & 0 & 0 & 0 & 0\cr 
               0 & 0 & 0 & y ^8& 0 & 0 & 0 & 0\cr 
               \bar{x}^8 & 0 & 0 & 0 & 0 & 0 & 0 & 0\cr 
               0 & \bar{y}^8 & 0 & 0 & 0 & 0 & 0 & 0\cr 
               0 & 0 & 0 & 0 & 0 & 0 & z^8 & 0\cr 
               0 & 0 & 0 & 0 & 0 & 0 & 0 & w^8\cr 
               0 & 0 & 0 & 0 & \bar{z}^8 & 0 & 0 & 0\cr
               0 & 0 & 0 & 0 & 0 & \bar{w}^8 & 0 & 0\cr }
\nonumber\\
X^9&=&\pmatrix{  0 & x^9 & 0 & 0 & 0 & 0 & 0 & 0\cr 
               \bar{x}^9 & 0 & 0 & 0 & 0 & 0 & 0 & 0\cr 
               0 & 0 & 0 & y^9 & 0 & 0 & 0 & 0\cr 
               0 & 0 & \bar{y}^9 & 0 & 0 & 0 & 0 & 0\cr 
               0 & 0 & 0 & 0 & 0 & z^9 & 0 & 0\cr 
               0 & 0 & 0 & 0 & \bar{z}^9 & 0 & 0 & 0\cr 
               0 & 0 & 0 & 0 & 0 & 0 & 0 & w^9\cr
               0 & 0 & 0 & 0 & 0 & 0 & \bar{w}^9 & 0\cr  }.\label{xs}
\eeqs}}

It is possible to write the action on the probe brane in manifest 
${\cal N} = (1,1)$ language by using the \emph{three-dimensional}
${\cal N}=1$ superfield notation described in \cite{Gates:1983nr}. 
By taking the superfields to be independent on the last space 
coordinate we get the ${\cal N} = (1,1)$ superfields. We will use 
the same three dimensional notation of \cite{Gates:1983nr} and define
a vector superfield ${\bf\Gamma}_\alpha$, containing the gauge potential
and the field $X^2$, coming from the third component of the three dimensional
gauge fields.
We also introduce the associated superfield strength
${\bf W}_\alpha = (1/2) D^\beta D_\alpha {\bf \Gamma}_\beta$ and
supercovariant derivative\footnote{Although our gauge theory is abelian, 
the couplings (charges) of the matter fields to the gauge fields come 
from reducing the maximally supersymmetric theory and thus can be conveniently
expressed as commutators by putting together the eight $U(1)$ gauge fields
into a diagonal matrix.}
$\nabla_\alpha = D_\alpha + i[ {\bf \Gamma}_\alpha, \cdot ]$.
The 7 scalar superfields 
${\bf X}^i$ parameterizing the motion in the orbifold directions have
projections: ${\bf X}^i| = X^i$,  
$\nabla_\alpha{\bf X}^i| = \psi^i_\alpha$
and  ${\bf \nabla^2 X}^i| = F^i$, 
where $F^i$ is the auxiliary field in ${\bf X}^i$.

The ${\cal N} = (1,1)$ Yang-Mills theory (sigma model) can then be written as
\beq
     S = \frac{1}{g^2}\int\;d\sigma^2\;d\theta^2 \mathrm{tr}
     \left({\bf W}^\alpha {\bf W}_\alpha
     -\frac{1}{2}\nabla^\alpha {\bf X}^i \nabla_\alpha {\bf X}^i -
      \frac{i}{3} \omega^{ijk} 
      {\bf X}^i {\bf X}^j {\bf X}^k \right), \label{action}
\eeq
where $\omega^{ijk}$ is the invariant 3rd-rank antisymmetric 
tensor of $G_2$, also known as the structure constant 
of the imaginary octonions. In our choice of indices for the coordinates,
running from $3$ to $9$, the non zero elements of $\omega$ are
\beq
    \omega^{789}=\omega^{569}=\omega^{468}=\omega^{394}=
    \omega^{358}=\omega^{367}=\omega^{475}=1, \label{ome}
\eeq
the other following by the total antisymmetry of $\omega$.

The very non trivial fact that makes the superpotential in (\ref{action})
work is that squaring the 7 F-terms
\beq
    F^i = i \omega^{ijk} X^j X^k,  \label{moment}
\eeq
one reproduces the usual quartic bosonic potential:
\beq
    \sum_{i} \mathrm{tr} \left(F^i F^i\right) =
    - \sum_{i<j} \mathrm{tr} \left([X^i,X^j]^2\right).
\eeq
Thus, the vanishing of the 7 F-terms is enough to guarantee the 
vanishing of all 21 relative commutators.

The 7-vector $F^i$ (\ref{moment})
is the analogue to our case of the moment map of \cite{Hitchin:1987ea}, 
\cite{Kronheimer:1989zs}! 
(We expand more on this at the end of next section).
We will refer to it as 
the ``octonionic'' moment map because of the presence of the structure constant
for the octonions.

It is amusing to notice that the same expression (\ref{action}) 
seems to hold
for the maximally supersymmetric case, where we drop the restrictions coming
from the orbifold projection and consider the superfields to be arbitrary
hermitian matrices. Thus, for the three dimensional ${\cal N} = 8$ 
theory, (which has
exactly the same form as (\ref{action}), only integrated over $d\sigma^3$),
we have a manifestly ${\cal N} = 1$ formulation in which the manifest
R-symmetry is $G_2$. 

\section{Probe analysis and the removal of the singularity
at the origin}

As usual, there are two possible branches that can be studied. The one
of interest here is the ``Higgs branch'', defined by setting the field
$X^2$ equal to a constant times 
the identity matrix. The other branch, known as the
``Coulomb branch'' is relevant, for instance, 
in the study of fractional branes.

The Higgs branch of the vacuum solution for the singular orbifold is given by 
the vanishing of the octonionic moment
map, 
\beq
     F^i=0 \label{singeq},
\eeq
which has solutions iff all the $X^i$ commute with each other.

It is not difficult to see that the vacuum equations imply
$|x^i|=|y^i|=|z^i|=|w^i|:=r^i$, so that we
can set $x^i=r^i e^{i\chi^i},\, y^i=r^i e^{i\psi^i},\, z^i=r^i e^{i\zeta^i},
\, w^i=r^i e^{i\omega^i}$. Plugging these expressions back
to the vacuum condition, we get 42 real equations (two from
each commutator)
for the 28 phases (there are 4 phases for each matrix $X^i$).
It would naively seem that the system is over-determined, but 
21 of the equations are redundant and we can solve in
terms of 7 appropriately chosen phases. It is straightforward
to check that a set of such phases is given by
$\{\chi^3,\, \psi^4,\, \zeta^5,\, \omega^6,\,
 \zeta^7,\, \omega^8,\, \omega^9  \}$.
The rest of the phases are then expressed as linear combinations
of the above: $\psi^3=-\omega^9+\psi^4,\, \zeta^3=-\omega^8+\zeta^5,\dots$
etc. On the other hand, under a gauge transformation 
parameterized by $\{\Lambda_1,\dots \Lambda_8\}$, 
$\{\chi^3,\, \psi^4,\, \zeta^5,\, \omega^6,\,
 \zeta^7,\, \omega^8,\, \omega^9  \}$
get shifted by $\{\Lambda_1-\Lambda_8,\dots , \Lambda_7-\Lambda_8  \}$
respectively. A careful analysis reveals that 
{\it all} the phases can be gauge-fixed to zero
modulo $\pi$. Consequently we are
left with a 7 dimensional moduli space parameterized by
$\pm r^i$. This is precisely the 
orbifold ${\bf R}^7/\Gamma$ we started with
and it is, of course, singular at the origin.
It is also straightforward to see, 
after integrating out the gauge field, that the metric is flat.

We are familiar with the fact that the coupling of the twisted metric moduli
to the brane modifies the bosonic potential in such a way that the 
supersymmetric vacuum is now smoothed out. This usually comes about via
D-term couplings, such as those generated by Fayet-Iliopoulos terms.
Here we face an immediate puzzle because in the model we are considering 
there are neither D-terms nor twisted metric moduli. 
In fact, looking at the expansion of 
${\bf \Gamma}_\alpha$ we see that the only scalar field is 
$B=D^\alpha {\bf \Gamma}_\alpha|$ which can be completely gauged away.
In the reduction to two dimensions, we have the possibility of constructing
a superfield $\mathbf{\Xi} 
= D_\alpha \gamma^2_{\alpha\beta}{\bf \Gamma}_\beta$,
which is gauge invariant ($\gamma^2_{\alpha\beta}$ is the last of the 
Dirac matrices in three dimensions that becomes the chirality matrix in two),
but that does not help either because its last component 
$F_{01} = D^2 \mathbf{\Xi}|$ is the electro-magnetic field strength and 
not an auxiliary field.

If there is a term that removes the singularity, it must come from a 
\emph{direct} coupling of the twisted fields 
$\phi_h$ ($h\in\{\alpha, \beta, \gamma, \alpha\beta, 
\alpha\gamma, \beta\gamma, \alpha\beta\gamma \}$)
to the matter superfields ${\bf X}^i$. There is one ``almost''
obvious candidate\footnote{One way to check this would be to perform an
explicit string calculation of the couplings $<\phi\phi XX>$ and
$<\phi X X X>$.},
the coupling of each twisted field 
$\phi_h$ to a ``twisted'' mass term 
$\mathrm{tr}\big({\bf X}^i{\bf X}^i\rho(h)\big)$. 
The part of this coupling that is relevant is
\beq
        S' = \frac{1}{g^2}\int\;d\sigma^2\;d\theta^2 \sum_{h\not=e}
      \phi_h\mathrm{tr}
     \left({\bf X}^i {\bf X}^i \rho(h) \right). \label{ra}  
\eeq

In the presence of this coupling, equation (\ref{singeq}) for the Higgs 
branch
is now given by
\beq
    F^i :=  i \omega^{ijk} X^j X^k-
\sum_{h\not=e}\phi_h \left\{X^i, \rho(h)\right\} =0 \label{smooth}
\eeq
$F^i$ being the modified (compared to (\ref{moment})) 
``octonionic'' moment map.

We now show that this modification removes the singularity 
at the origin. First of all, notice that, contrary to what happens with the 
usual D-terms, here the origin is always a solution to (\ref{smooth}). 
However,
we shall see that for generic values of $\phi_h$ the origin is an isolated 
point.

The first thing to notice about (\ref{smooth}) is that each choice of $i$
gives a set of four complex equations and their complex conjugates, 
equivalently, eight real equations. What does the trick is that, because of
the particular structure of the non zero elements of $\omega^{ijk}$ in
(\ref{ome}), the commutator $[X^j, X^k]$ has 
non-zero entries at the same places as $X^i$ does
(cf. (\ref{xs})). The form of $\omega^{ijk}$ is of course 
dictated by invariance under the orbifold group $\Gamma$. 

A second important property is that for each $\rho(h)$ there are
four $X$'s that anti-commute with it and three that commute.
In particular, the coupling in (\ref{ra}) is non zero only between $\phi_h$
and those superfields that are \emph{not} twisted by $R(h)$. 
However, one can easily see that it is possible to choose $\phi_h$ in such
a way that each of the $7\times 4=28$ complex fields making up the $X$'s
will appear linearly in one of the (\ref{smooth}). 
(In fact three non zero $\phi$'s 
are enough). 

Assuming that we have done as above and turned on a ``mass'' term for each
of the $28$ fields we see that the origin must be an isolated point. 
In fact, near the origin, the quadratic piece can be neglected and the 
equations simply impose $X^i\equiv 0$. What is more difficult is to show 
that there is a smooth solution away from the origin. This can be done to
first order in perturbation theory. Let us collectively 
denote by $t^A$, $A=1, ... 56$ the
real variables appearing in (\ref{smooth}).
For instance, one could set $x^3 = t^1 + i t^2$, $y^3 = t^3 + i t^4$,
... $w^9 = t^{55} + i t^{56}$. 
The superpotential becomes
\beq
    W[t] = \frac{1}{6} \sum_{ABC} Q^{ABC} t^A t^B t^C + 
           \frac{1}{2}\sum_A m^A t^A t^A,
\eeq
where the totally symmetric tensor $Q$ is 
determined by $\omega$ and the ``mass'' terms $m^A$ are just 
linear combinations of $\phi$'s.
Equations (\ref{smooth}) can then be rewritten as a set of 56 real equations
\beq
    \frac{1}{2}\sum_{BC} Q^{ABC} t^B t^C  =  m^A t^A,
\eeq
Let us write $t^A = \hat{t}^A + \tau^A$
where $\hat{t}^A$ is a solution of the undeformed equations
\beq
   \frac{1}{2}\sum_{BC} Q^{ABC} \hat{t}^B \hat{t}^C  = 0,    \label{unde}
\eeq
and $\tau^A$ is a perturbation. 
Moreover, 
let us assume 
that all $|m^A/\hat{t}^A|,\, |\tau^A/\hat{t}^A|<<1$.
Expanding to first order
we obtain an inhomogeneous linear equation for the $\tau$'s
\beq
    \sum_{BC}  Q^{ABC} \hat{t}^B \tau^C  =  m^A \hat{t}^A.
\eeq
We require that the system have a 14 dimensional space of solutions,
corresponding to the 7 coordinates of the manifold and 7 gauge directions.
We have performed the computation using Mathematica and found that it does!
(Not surprisingly, one can check that the addition of 
an eighth ``untwisted'' mass 
term $\sim \mathrm{tr}\left({\bf X}^i {\bf X}^i\right)$ to the potential
would lift the Higgs branch completely).

The perturbation theory we used is good far from the origin whereas near
the origin we have used another approximation to show 
that the origin is 
an isolated point. This is what can be shown from this rather 
general analysis. What has {\it not} been shown 
is that the new, perturbed, solution is smooth everywhere.
Our arguments do not rule out the possibility that
a new (different from the origin) singularity develops.
The most direct way to settle this would be to obtain
the exact solution to the perturbed equation. We hope to
report on this in some future work.

Let us now
compare to the hyperK\"{a}hler quotient construction 
\cite{hit}, \cite{Hitchin:1987ea},
\cite{Kronheimer:1989zs}.
The ingredients are: a large parent space ${\cal M}$,
a gauge group $G$, a Lie-algebra-valued triplet ${  \mu^i,\, i=1,2,3}$
and a triplet of numbers $\zeta^{i}$ valued in the center 
${\cal Z}$ of $G/U(1)$. I.e.
\beqs
{  \mu}^i: &{\cal M}&\rightarrow {\bf R}^3\otimes {\cal L}(G/U(1))\nonumber\\
&\zeta^{i}&\in {\bf R}^3\otimes {\cal Z}
\eeqs
The hyperK\"{a}hler quotient is then simply  
${\bf  \mu}^{-1}(\zeta^{i})/(G/U(1))$.

These objects fit into  a nice physical
picture \cite{Douglas:1996sw}: $\cal M$ is the space parameterized
by the hypermultiplets of the low-energy effective theory on the probe,
$G$ is the gauge group of the latter, ${\bf  \mu}$ are the
$D$-terms of the theory and $\zeta^{i}$ 
are the Fayet-Iliopoulos terms. The hyperK\"{a}hler quotient in
this language is the moduli space of vacua of
the theory on the probe, i.e. the set of 
gauge-invariant solutions to the
D-flatness conditions.

In our case the analogue
of the FI terms are the 7 real moduli $\phi_h$
and the analogue of
the moment map is the, now $\phi_h$-dependent,
octonionic map of (\ref{smooth}). We have,
\beqs
{\bf F}(\phi): 
&{\cal M}&\rightarrow {\bf R^7}\otimes {\cal L}(G/U(1))\nonumber\\
&\phi_h&\in {\bf R^7}\otimes {\cal Z}
\eeqs
The modified 
vacuum moduli space 
is simply $\mathit{Ker}({\bf F}(\phi))/(G/U(1))$.

\section{Conclusions}

We have seen that D-branes can be useful tools 
in the  study of $G_2$ orbifolds and
proposed that they remove 
orbifold singularities in a novel way -- by a twisted
mass term instead of the D-terms.
In the process of doing so, we have come across a 
proposal for a generalization of the
moment map.

Clearly, our construction generalizes to many other situations and 
there are various possible directions along these lines. An understanding of
the possible discrete subgroups of $G_2$ would tell us which are the
interesting cases to be studied. A better
understanding of the geometry of these orbifolds is needed, perhaps even
explicit metrics can be constructed this way. 

Unfortunately we have not been able to prove that our proposed
bulk coupling renders the vacuum moduli space smooth everywhere. The most
direct way to show this would be to obtain the exact metric on the
modified vacuum moduli space.
What we have shown is that a) the singularity at the origin
is removed (becomes an isolated point) and b) the first-order 
perturbation does not lift the space of vacua.

It would clearly be interesting 
to justify (or rule out!) the presence of the twisted mass term by a 
direct string world-sheet calculation.

\section*{Acknowledgments}

We would like to thank L. Brink, M. Cederwall, M. Henningson 
and B. E. W. Nilsson for discussions.
We would also like to 
thank G. Moore for email correspondence and B. Acharya for pointing out 
an erroneous interpretation of twisted moduli in the previous version
of this paper.
This work is supported in part by the European Union RTN contracts
HPRN-CT-2000-00122.

\end{document}